\documentclass[conference]{IEEEtran}

  	\usepackage[pdftex]{graphicx}
  	\graphicspath{{../pdf/}{../jpeg/}}
	\DeclareGraphicsExtensions{.pdf,.jpeg,.png}

	\usepackage[cmex10]{amsmath}
    \usepackage{amssymb}
	\usepackage{algorithmic}
	\usepackage{array}
	\usepackage{mdwmath}
	\usepackage{mdwtab}
	\usepackage{eqparbox}
	\usepackage{url}
    \usepackage{cite}
    \usepackage{color}
    \usepackage{comment}
   \usepackage[compact]{titlesec} 
   \usepackage{bm}
\titlespacing*{\section}{0pt}{*2}{*0.7}
\titlespacing*{\subsection}{0pt}{*1.4}{*0.6}
\titlespacing*{\subsubsection}{0pt}{*1.1}{*0.5}

\usepackage[colorlinks=true,citecolor=blue,linkcolor=blue,urlcolor=blue]{hyperref}
\renewcommand{\abstractname}{}
\begin{document}

\title{\LARGE Structure-resolved free energy estimation of the 38-atom Lennard--Jones cluster via population annealing}

\author{\authorblockN{Akie Kowaguchi\authorrefmark{1} and  Koji Hukushima\authorrefmark{2} }

\authorblockA{\authorrefmark{1}Department of Pharmacology, School of Medicine, Keio University, Tokyo, Japan}

\authorblockA{\authorrefmark{2}Graduate School of Arts and Sciences, The University of Tokyo, Tokyo, Japan}}

\pagenumbering{arabic} 
\thispagestyle{plain}  
\pagestyle{plain}      
\makeatletter
\twocolumn[
\begin{@twocolumnfalse}
\maketitle

\vspace{-1.5em}
\begin{center}
{\small \today}
\end{center}
\vspace{-1.5em} 

\begin{abstract}
We systematically investigate the thermodynamic landscape of the 38-atom Lennard--Jones cluster LJ$_{38}$ using Population Annealing (PA), a method suited for systems with challenging double-funnel energy landscapes. By employing an adaptive temperature schedule, we demonstrate that thermodynamic observables, such as internal energy and heat capacity, converge robustly when the population size is sufficiently large. To gain deeper insights into the competing basins, we introduce an integrated framework that combines PA reweighting factors with structure-resolved analysis. Using quenched configurations characterized by potential energy and Steinhardt's bond-orientational order parameters, we identify three structural basins, FCC-like, icosahedral, and liquid-like, via dimensionality reduction and clustering. This framework enables the direct computation of structure-resolved free energy differences from population fractions, providing a quantitative mapping of the thermodynamic competition between the funnels. The resulting structural crossovers are consistent with the heat-capacity peak, demonstrating PA as a promising and scalable framework for structure-resolved thermodynamics in complex molecular systems. 
\end{abstract}

\vspace{0.5em} 
\end{@twocolumnfalse}
]

\IEEEoverridecommandlockouts

\IEEEpeerreviewmaketitle


\section{Introduction}

Achieving accurate equilibrium sampling in systems with complex energy landscapes remains a fundamental challenge in molecular simulation. In such systems, conventional Markov chain Monte Carlo (MCMC) and molecular dynamics (MD) methods often suffer from broken ergodicity, in which trajectories become trapped in metastable states for timescales far exceeding the feasible simulation times. While modern supercomputing architectures provide massive parallelism, many widely used enhanced-sampling methods require careful tuning to maintain efficiency as the computation scale increases. These considerations motivate the development of sampling approaches that can both efficiently equilibrate and yield reliable thermodynamic quantities, including free energy estimates, while scaling seamlessly to massively parallel platforms.

In this study, we focus on Population annealing (PA) \cite{hukushima2003population}, a sequential Monte Carlo method that has proven effective for a wide range of frustrated systems. Initially developed for discrete spin systems~\cite{hukushima2003population, Wang_2015}, the efficiency of PA has recently been demonstrated in continuous systems, including binary hard-sphere mixtures\cite{Callaham_2017} and some molecular models~\cite{Christiansen_2019PRL}. For equilibrating complex landscapes~\cite{Machta_2010} and searching for ground states~\cite{Wang_2015_Comp}, PA exhibits a level of performance comparable to parallel tempering (replica exchange method)~\cite{hukushima1996exchange}. However, PA has distinct practical advantages, including its inherent scalability for massively parallel architectures~\cite{Weigel_2017, BARASH2017341} and its ability to provide direct free energy estimates. Although PA does not fundamentally accelerate the local mixing of its underlying MCMC kernels, its primary strength lies in exploiting the statistical power of a large population to overcome high free energy barriers. The multi-point exploration maintains ensemble diversity and reduces weight imbalances, enabling sampling of multimodal landscapes that are otherwise poorly explored by conventional trajectories.

Atomic clusters provide a compact, yet stringent framework for evaluating such sampling algorithms. Despite their modest computational cost, they present demanding equilibration and sampling problems. In cluster science, competing structural motifs reflect the general principles of stability and packing in diverse systems, ranging from quasicrystals~\cite{shechtman1984metallic,tsai1987stable,ebert} to metal clusters~\cite{marks1981425,vanderVelden1981} and self-assembled nanostructures~\cite{fan2010self,de2015entropy}.
For simple pair-potential systems, the structural preference is determined by a competition between reducing the number of surface atoms and minimizing the strain in interatomic distances. Lennard--Jones (LJ) clusters therefore serve as widely used model systems for studying competing motifs and size-dependent structural transitions\cite{noya2006structural,doye1999evolution,doye2002entropic,northby1987structure,sehgal2014phase}.

\begin{figure}[t]
\centering
\includegraphics[width=3.5in]{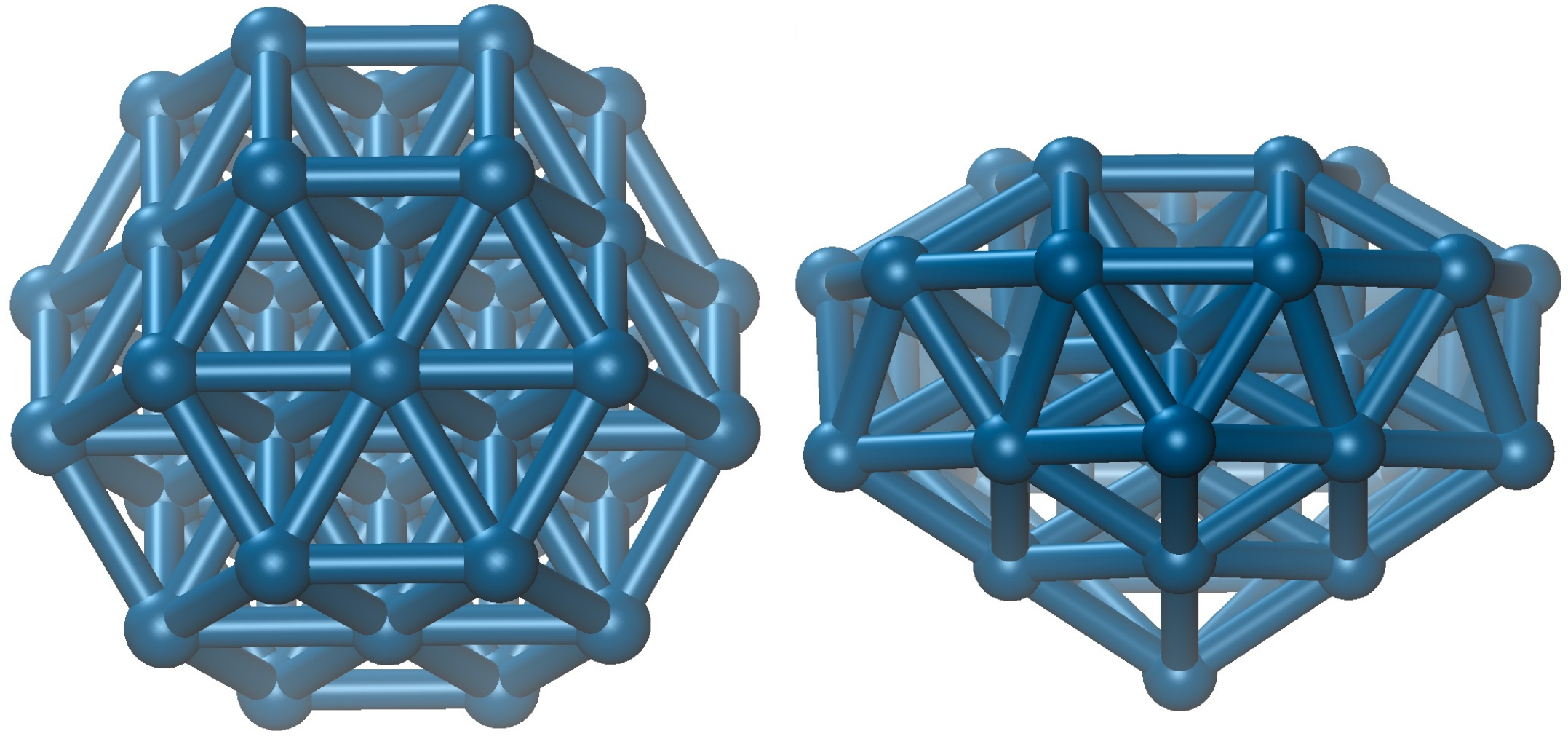}
\caption{Representative low-energy structures of LJ$_{38}$: the FCC truncated-octahedral global minimum (left) and a competing icosahedral local minimum (right). Despite having similar energies, the funnels are separated by a substantial free energy barrier, making LJ$_{38}$ a demanding test case for sampling algorithms.}
\label{fig:lj38_motifs}
\end{figure}

Among these, the 38-atom Lennard--Jones cluster (LJ$_{38}$) is notoriously difficult to equilibrate due to its pronounced double-funnel landscape. As shown in Fig.\ref{fig:lj38_motifs}, the global minimum is a face-centered cubic (FCC) truncated octahedron, which competes with an icosahedral funnel based on an incomplete Mackay icosahedron. Doye and Wales \cite{doye1995effect,doye1997structural} highlighted a trade-off between surface coordination and interior strain: icosahedral motifs achieve higher nearest-neighbor coordination for a given surface area, whereas fcc packings favor a lower-strain interior, with fewer surface contacts.
Near the solid–solid crossover, the two funnels are thermodynamically competitive, but the barrier between them remains prohibitively high for thermal fluctuations to induce frequent transitions.

Due to these features, LJ$_{38}$ has long served as a challenging benchmark for the evaluation of extended ensemble methods\cite{doye1999double,partay2010efficient,poulain2006performances,bogdan2006equilibrium,lv2012particle,goedecker2004minima,wales2006potential,neirotti2000phase,avendano2016firefly,predescu2005thermodynamics,schaefer2014minima,kanayama2023structure,sharapov2007solid,liu2005convergence}, including replica exchange method\cite{hukushima1996exchange}, multicanonical sampling\cite{berg1991multicanonical,berg1992multicanonical}, Wang–Landau sampling\cite{wang2001efficient}, basin hopping\cite{wales1997global}, nested sampling\cite{skilling2004nested} and related advanced techniques, each possessing inherent strengths and limitations.

The replica exchange method is a widely used and easy-to-implement strategy for enhanced sampling and has proven highly effective in many molecular simulations. At the same time, its efficiency can be limited by the need to design appropriate temperature intervals that maintain reasonable exchange probabilities. In massively parallel settings, simply increasing the number of replicas to maintain exchange probabilities may lead to overly dense temperature spacings, which do not necessarily improve global mixing. For LJ$_{38}$, previous studies have also noted that simulations initialized from randomized configurations may struggle to reach the FCC truncated-octahedral global minimum \cite{liu2005convergence, cezar2017parallel}.

Wang–Landau and multicanonical methods aim to flatten the energy distribution to promote barrier crossing; however, for continuous cluster systems, energy binning and the fact that energy alone is often an incomplete reaction coordinate can still lead to slow mixing between distinct basins. Indeed, one-dimensional Wang–Landau simulations using energy as a single variable have been reported to fail in reproducing the equilibrium between the octahedral and the entropically favored icosahedral funnels\cite{poulain2006performances}.
Including an additional order parameter was shown to improve mixing, but it requires system-specific choices and adds practical complexity.
Nested sampling enables direct calculation of free energy and is highly parallelizable. At each iteration, the method assumes approximately uniform sampling within the configuration space constrained below a given energy threshold. However, for systems with highly complex or rugged landscapes, maintaining this uniformity can be challenging, which may make it difficult to fully capture all relevant low-energy basins and to accurately estimate free energy differences.

Motivated by the inherent scalability of PA and its capability for direct free energy estimation, we demonstrate its utility for structure-resolved thermodynamics using the LJ$_{38}$ cluster as a benchmark. While we confirm the convergent behavior of thermodynamic observables as a function of population size and quantify sampling stability using an adaptive temperature schedule~\cite{Doucet2001,adaptive_resampling}, our main focus is to present a systematic workflow that integrates PA reweighting with the analysis of configurations sampled by PA and subsequently quenched to their local minima. This approach enables us to quantitatively obtain structure-resolved free energy differences and characterize the temperature-dependent competition among the FCC-like, icosahedral, and liquid-like basins without relying on traditional harmonic approximations. The resulting PA-based framework provides a practical route to mapping finite-temperature free energy landscapes in molecular systems with competing motifs.

This paper is organized as follows. Section~\ref{sec:method} describes PA, the adaptive temperature schedule, and the structural descriptors used for classification. Section~\ref{sec:results} presents the simulation results, including convergence of thermodynamic observables and structure-resolved free energy profiles based on dimensionality reduction and clustering. Section~\ref{sec:conclusion} summarizes the main findings and outlines directions for future work.

\section{Method and Model}
\label{sec:method}
We first describe PA and the MCMC-based local update scheme used for sampling. This is followed by the definition of the Lennard--Jones cluster model and our classification protocol, which combines energy minimization and clustering to identify structural basins and quantify their relative stability at finite temperature.

\subsection{Population Annealing}
Consider a cluster of $N$ particles in three-dimensional space. 
We denote a specific configuration by $\bm{X} \in \mathbb{R}^{3N}$, representing the coordinates of all particles.  For a given potential energy $E(\bm{X})$, the canonical distribution at inverse temperature $\beta=(k_\text{B}T)^{-1}$ is defined as 
\begin{equation}
\mathcal{P}_\beta(\bm{X})=\frac{e^{-\beta E(\bm{X})}}{Z(\beta)},
\end{equation}
where $Z(\beta)$ is the configurational partition function given by the integral over all possible configurations: 
\begin{equation}
Z(\beta)=\int e^{-\beta E(\bm{X})}\,\mathrm{d}\bm{X}. 
\end{equation}
In this work, we focus on the configurational part of the phase space, as the kinetic contributions to the partition function are analytically integrable and do not affect the relative structural stability. 

\begin{figure}[] 
\centering
\includegraphics[width=2.5in]{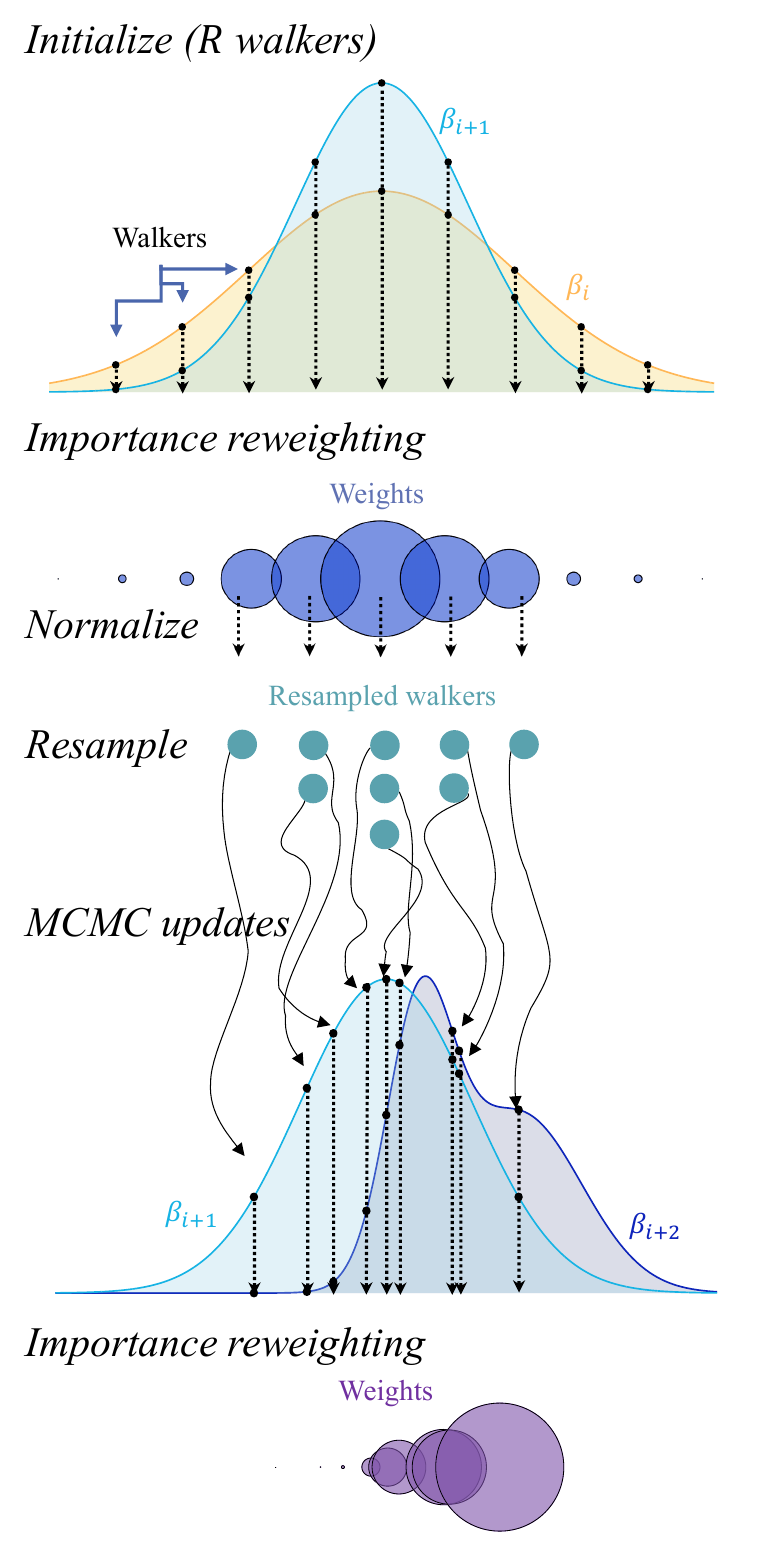}
\caption{Schematic overview of the population annealing procedure (top to bottom). Walkers equilibrated at inverse temperature $\beta_i$ are reweighted to the next inverse temperature $\beta_{i+1}$, followed by a systematic resampling process to maintain a constant population size. Subsequently, short MCMC updates are applied to decorrelate the walkers and enhance configurational diversity at $\beta_{i+1}$. The bottom of the schematic presents the overlap between the energy distributions at $\beta_{i+1}$ and $\beta_{i+2}$, which determines the importance weights for the next step. Circle size indicates the relative weights, and the bell-shaped curves schematically represent the energy distributions at each temperature. The procedure is repeated along the annealing schedule.}
\label{PAdiagram}
\end{figure}

PA is a sequential Monte Carlo method that generates samples from a target distribution by evolving a population of $R$ independent walkers. Each walker has a configuration $\bm{X}$ and is initially equilibrated at a sufficiently high temperature $\beta_0$. It is then cooled according to an inverse–temperature sequence
\begin{equation}
\{\beta_i\}\qquad i=0,1,\dots,L,
\end{equation}
where $\beta$ increases from a high–temperature value $\beta_0$ to the low–temperature target $\beta_L$. 
We denote the step size by $\Delta\beta_i=\beta_{i+1}-\beta_i>0$. At each annealing stage (or temperature step), the walkers undergo two main processes: short MCMC updates to explore the local phase space, and resampling based on importance weights to maintain equilibrium statistics as $\beta$ increases. Fig.\ref{PAdiagram} provides a schematic overview of a single temperature step in the PA procedure.

\subsubsection{Resampling and Weight Update}
\label{subsec:resampling}
In our implementation of PA, the population size is kept fixed at $R$ throughout the annealing process. While some variants of PA allow $R$ to fluctuate, keeping $R$ constant makes load balancing easier in parallel computing environments and simplifies the tracking of ancestral lineages. 

Let $\{\bm{X}_j\}_{j=1}^{R}$ be the configurations of $R$ walkers sampled from $\mathcal{P}_{\beta_i}$ at inverse temperature $\beta_i$, with corresponding potential energies $E_j=E(\bm{X}_j)$. At the initial high temperature $\beta_0$, the population is set to be unweighted, i.e., each walker has an identical weight $w_j^0=1$. 
To evolve the population to the next temperature $\beta_{i+1}=\beta_i+\Delta\beta_i$, each walker is assigned an importance weight $w_j^{i+1}$ defined as  
\begin{equation}
w_j^{i+1} \;=\; w_j^{i} \frac{\exp\!\left(-\Delta\beta_iE_j\right)}{Q_i},
\qquad j=1,\dots,R,
\end{equation}
where the normalization factor $Q_i$ is given by 
\begin{equation}
Q_i \;=\; \frac{1}{R}\sum_{j=1}^{R}w_j^{i}\exp\!\left(-\Delta\beta_iE_j\right).
\end{equation}
The population is then updated by resampling $R$ new walkers from the current set, where the expected number of copies for walker $j$ is proportional to $w_j^{i+1}$. Crucially, this resampling procedure resets the importance weights of the new population to unity. This ensures that walkers entering the subsequent MCMC updates and the next reweighting step form an unweighted ensemble at ${\beta_{i+1}}$, avoiding the accumulation of weight variance across multiple temperature steps. 
Among several resampling schemes~\cite{Tutorial_particle_filtering}, we adopt \emph{systematic resampling}, which in our simulations yielded lower variance in the resulting estimates than multinomial resampling. This is consistent with previous comparative studies of resampling schemes in PA~\cite{gessert2023resampling}.

\subsubsection{Adaptive Temperature Scheduling}
Theoretically, PA is an exact sampling method in the infinite-population limit $(R\to\infty)$ for any annealing schedule. In practice, the performance and finite-$R$ bias depend strongly on the choice of the temperature increments $\Delta\beta_i$. To maintain sufficient population diversity, we adaptively determine $\Delta\beta_i$ such that the effective sample size (ESS) approaches a target value~\cite{Doucet2001, adaptive_resampling}. The ESS at the transition to $\beta_{i+1}$ is defined as
\begin{equation}
\mathrm{ESS}_{i+1} \;=\; \frac{\left(\sum_{j=1}^{R} w_j^{i+1}\right)^2}{\sum_{j=1}^{R} (w_j^{i+1})^2}, 
\end{equation}
which takes values from $1$ to $R$. A value of $\mathrm{ESS}_{i+1} \simeq R$ indicates nearly equal weights, whereas $\mathrm{ESS}_{i+1}\simeq 1$ implies that a few walkers dominate the population, causing a rapid loss of ancestral diversity. While conventional adaptation schemes often rely on the coefficient of variation of weight factors~\cite{Christiansen_2019}, we adaptively choose $\Delta\beta_i$ by directly controlling the ESS to maintain a robust representation of configuration space. Specifically, $\Delta\beta_i$ is selected so that $\mathrm{ESS}_{i+1}$ stays above a prescribed threshold, which helps preserve lineage diversity during resampling.

\subsubsection{Estimation of Observables and Free Energy}
A key feature of PA is its ability to maintain an equilibrated ensemble that follows the Boltzmann distribution at each inverse temperature $\beta_i$ during cooling. In contrast to simulated annealing~\cite{SA1983, Kirkpatrick1984}, which is primarily an optimization tool for finding ground states, PA provides a sequence of equilibrium samples, enabling the calculation of thermal averages at any intermediate temperature. 

As discussed in the previous subsection, the resampling step resets the importance weights of all the walkers to unity. Consequently, the expectation value of a physical observable $\mathcal{A}$ at $\beta_i$ is simply estimated by the sample mean over the population: 
\begin{equation}
    \langle\mathcal{A}\rangle_{\beta_i} \approx\frac{1}{R}\sum_{j=1}^{R}\mathcal{A}(\bm{X}_j). 
\end{equation}
This allows for the straightforward measurement of thermodynamic quantities and structural order parameters as the temperature is lowered, using the same estimators as in standard MCMC. 

In addition to ensemble averages, PA enables the estimation of the partition function and the total free energy, which are often challenging to obtain using conventional MCMC. The ratio of the partition functions at consecutive temperature steps is given by the normalization factors $Q_i$. By accumulating these factors, the total free energy change from $\beta_0$ to $\beta_L$ is estimated as
\begin{equation}
\beta_L F(\beta_L)-\beta_0 F(\beta_0) \;=\; -\sum_{k=0}^{L-1}\ln Q_k. 
\end{equation}
Using a high-temperature reference state in the simulation cell, this relation yields the free energy change along the schedule down to low temperatures. For Lennard--Jones clusters, such free energy differences are essential for quantifying the relative stability of competing structural basins at finite temperature.

\subsubsection{Metropolis--Adjusted Langevin Algorithm}
To efficiently sample the configurations during the short MCMC updates at each inverse temperature $\beta_i$, we employ the Metropolis--Adjusted Langevin Algorithm (MALA)~\cite{Rossky_1978,Roberts1996ExponentialCO}. In contrast to conventional local Metropolis updates, MALA incorporates the gradient of the target log-density, defined as the potential energy scaled by temperature, to guide the walkers towards regions of high probability. This gradient-based approach has been shown to be significantly more efficient than the standard random-walk Metropolis method in high-dimensional systems~\cite{Roberts_2002}, making it practically suitable for the complex energy landscapes of Lennard--Jones clusters. 

For a configuration $\bm{X}\in\mathbb{R}^{d}$ with $d=3N$ in three-dimensional space, the gradient of the log-density under the canonical distribution is given by 
\begin{equation}
\nabla \log \mathcal{P}_{\beta_i}(\bm{X}) \;=\; - \beta_i \nabla E(\bm{X}).
\end{equation}
A new configuration $\bm{Y}$ is proposed based on the Euler--Maruyama discretization of overdamped Langevin dynamics: 
\begin{equation}
\bm{Y} \;=\; \bm{X} \;-\; \frac{h^2 \beta_i}{2} \nabla E(\bm{X}) \;+\; h \,\bm{\xi},
\end{equation}
where $h>0$ is the integration step size, and $\bm{\xi}\sim\mathcal{N}(0,\bm{I})$ represents Gaussian noise with $\bm{I}$ being the $d\times d$ identity matrix. The corresponding Gaussian proposal density $q(\bm{X}\to\bm{Y})$ is defined as
\begin{equation}
q(\bm{X} \to \bm{Y}) \;=\; \mathcal{N}\!\left(\bm{Y} \,;\, \bm{X} - \frac{h^2 \beta_i}{2} \nabla E(\bm{X}), \, h^2 \bm{I} \right).
\end{equation}
To ensure the detailed balance condition, the proposed move is accepted with the Metropolis--Hastings probability:
\begin{equation}
\alpha(\bm{X},\bm{Y}) \;=\; \min\left( 1,\;
\frac{\mathcal{P}_{\beta_i}(\bm{Y})\, q(\bm{Y} \to \bm{X})}{\mathcal{P}_{\beta_i}(\bm{X})\, q(\bm{X} \to \bm{Y})}
\right).
\end{equation}
In our simulations, the step size $h$ is adjusted to maintain the acceptance probability at a certain value.

\subsection{Lennard--Jones Clusters}

We consider a cluster of $N$ particles with positions $\{\mathbf r_i\}_{i=1}^{N}$ interacting via the Lennard--Jones (LJ) pair potential.
The distance between particles $i$ and $j$ is denoted as
$
r_{ij} = \left| \mathbf r_i - \mathbf r_j \right|
$,
and the LJ pairwise interaction is given by 
\begin{equation}
V(r_{ij}) = 4\varepsilon \left[ \left( \frac{\sigma}{r_{ij}} \right)^{12} - \left( \frac{\sigma}{r_{ij}} \right)^{6} \right].
\end{equation}
The total potential energy of a configuration $\bm{X}=\{\bm{r}_i\}$ is the sum over all pairs: 
\begin{equation}
E(\bm{X})=\sum_{i<j} V(r_{ij}). 
\end{equation}
In this work, we focus on the $N=38$ cluster. 
Throughout our analysis, the internal energy is defined as the canonical average of the potential energy, $U(\beta)=\langle E\rangle_{\beta}$, and we report $U^*=U/\varepsilon$.
Unless otherwise stated, we use reduced LJ units with $\varepsilon=\sigma=1$ and do not apply a potential cutoff, accounting for all pairwise interactions within the cluster. Simulations are performed in a cubic periodic cell chosen sufficiently large that interactions with periodic images are negligible.

To characterize the structural order of the cluster, we employ the Steinhardt order parameters\cite{steinhardt1983bond}. A bond is defined between particles $i$ and $j$ when their separation satisfies $r_{ij}\le r_c$, where we set $r_c=1.5\sigma$. For each bond, we define the unit bond vector by
$\hat{\mathbf r}_{ij}=(\mathbf r_j-\mathbf r_i)/r_{ij}$.
For a given degree $l$, the global bond-orientational order is quantified by  the averaged spherical-harmonic components: 
\begin{equation}
\bar q_{lm}=\frac{1}{N_{\mathrm b}}\sum_{i<j} \Theta(r_c-r_{ij}) Y_{lm}(\hat{\mathbf r}_{ij}),
\qquad m=-l,\dots,l,
\end{equation}
where $Y_{lm}$ denotes the spherical harmonics of degree $l$ and order $m$, $N_{\mathrm b}$ is the total number of identified bonds, and $\Theta$ is the Heaviside step function. 
The corresponding rotationally invariant order parameter is defined as 
\begin{equation}
Q_l
= \left(\frac{4\pi}{2l+1}\sum_{m=-l}^{l}\left|\bar q_{lm}\right|^2\right)^{1/2}.
\end{equation}
Although $Q_4$ and $Q_6$ are commonly used to distinguish close-packed crystalline and icosahedral symmetries, we compute $Q_l$ for $l=2,\dots,12$ to provide a comprehensive structural characterization without a priori assumptions.

\subsection{Structural Classification and Free Energy Differences}
\label{subsec:free_energy}
While PA enables the direct estimation of the total free energy, a detailed understanding of the LJ$_{38}$ cluster requires connecting these thermodynamic quantities to specific structural motifs. Unlike optimization methods such as Basin Hopping~\cite{wales1997global}, PA provides equilibrium distributions at finite temperatures. To quantitatively evaluate the relative stability of competing structures, such as the icosahedral and truncated octahedral states, the configuration space must be partitioned into distinct basins, and their respective free energies must be computed. 

For this purpose, we employ a multi-step classification procedure. 
First, each configuration sampled during the PA run is quenched to its nearest local minima using the Fast Inertial Relaxation Engine (FIRE) algorithm~\cite{bitzek2006structural}. This mapping to \textit{inherent structures} removes thermal noise, allowing for a clearer identification of the underlying structural symmetry. 
Each quenched configuration is then characterized by a feature vector consisting of its reduced internal energy $E^*$ and the bond-orientational order parameters $Q_2, \ldots, Q_{12}$. 
To identify the most informative structural descriptors without a priori bias, we apply Principal Component Analysis (PCA) to standardized feature vectors. This step allows us to determine whether higher-order $Q_l$ parameters provide additional resolution for distinguishing structural motifs beyond the standard $Q_4$ and $Q_6$ metrics. We then perform $k$-means clustering\cite{macqueen1967multivariate} in the PCA-reduced space to group the quenched configurations into distinct structural families.
For the LJ$_{38}$ cluster, we set the number of clusters to $k=3$, which corresponds to the primary structural basins: the FCC-like, the icosahedral, and the liquid-like states.

Following the classification of all walkers at each temperature step, the structure-resolved free energies $F_c^*(T^*)$ for each structural family $c$ can be accurately evaluated. As discussed in Sec.~\ref{subsec:resampling}, the resampling step resets the importance weights of all walkers to unity. Consequently, the equilibrium contribution of each structural family $c$ is directly proportional to its population count $R_c$. 
By combining this with the total ensemble free energy $F^*_\text{tot}(T^*)$ obtained from the PA reweighting and normalization factors, the free energy difference of each structural family $c$ is estimated by   
\begin{equation}
    F_c^*(T^*) = F^*_\text{tot}(T^*) - T^*\ln \left(\frac{R_c(T^*)}{R}\right),
\end{equation}
where $R$ is the total population size. Similarly, the free energy difference between any two basins, $c$ and $c'$, is straightforwardly obtained from their population ratio, 
\begin{equation}
\Delta F_{c\leftarrow c'}^*(T^*)
= -T^*\ln\left(\frac{R_c(T^*)}{R_{c'}(T^*)}\right).
\end{equation}
This approach provides a framework for analyzing the temperature-dependent relative stability of emergent structural phases and their contributions to the overall thermodynamic landscape. 

\section{Numerical Results}
\label{sec:results}
\subsection{Simulation Setup}
We performed PA simulations of the LJ$_{38}$ cluster using an inverse-temperature schedule from $\beta_0=1.0$ to $\beta_L=20.0$.
The adaptive schedule based on the effective sample size (ESS) criterion described above yielded approximately 140 temperature points.
At each step $\beta_i\!\to\!\beta_{i+1}$, the increment $\Delta\beta_i=\beta_{i+1}-\beta_i>0$ was chosen by binary search to satisfy $\mathrm{ESS}_{i+1}/R=0.95$ within a tolerance of $0.01$.
Numerical stability during weight evaluation at low temperatures was ensured by employing a stabilized log-sum-exp scheme.

Following each resampling step, each walker performed $2.0\times 10^{4}$ MALA steps. The MALA step size $h$ was tuned at each temperature to maintain a target acceptance rate of approximately $0.5$. To evaluate the convergence of our results, we varied the population size $R=1000,\,2000,\,4000,\,8000,$ and $16000$. 
For each $R$, we conducted $10$ independent PA runs with distinct random seeds. Thermodynamic observables are averaged over independent runs with standard-error bars.

\subsection{Thermodynamic Observables}

\begin{figure}[t] 
\centering
\includegraphics[width=3in]{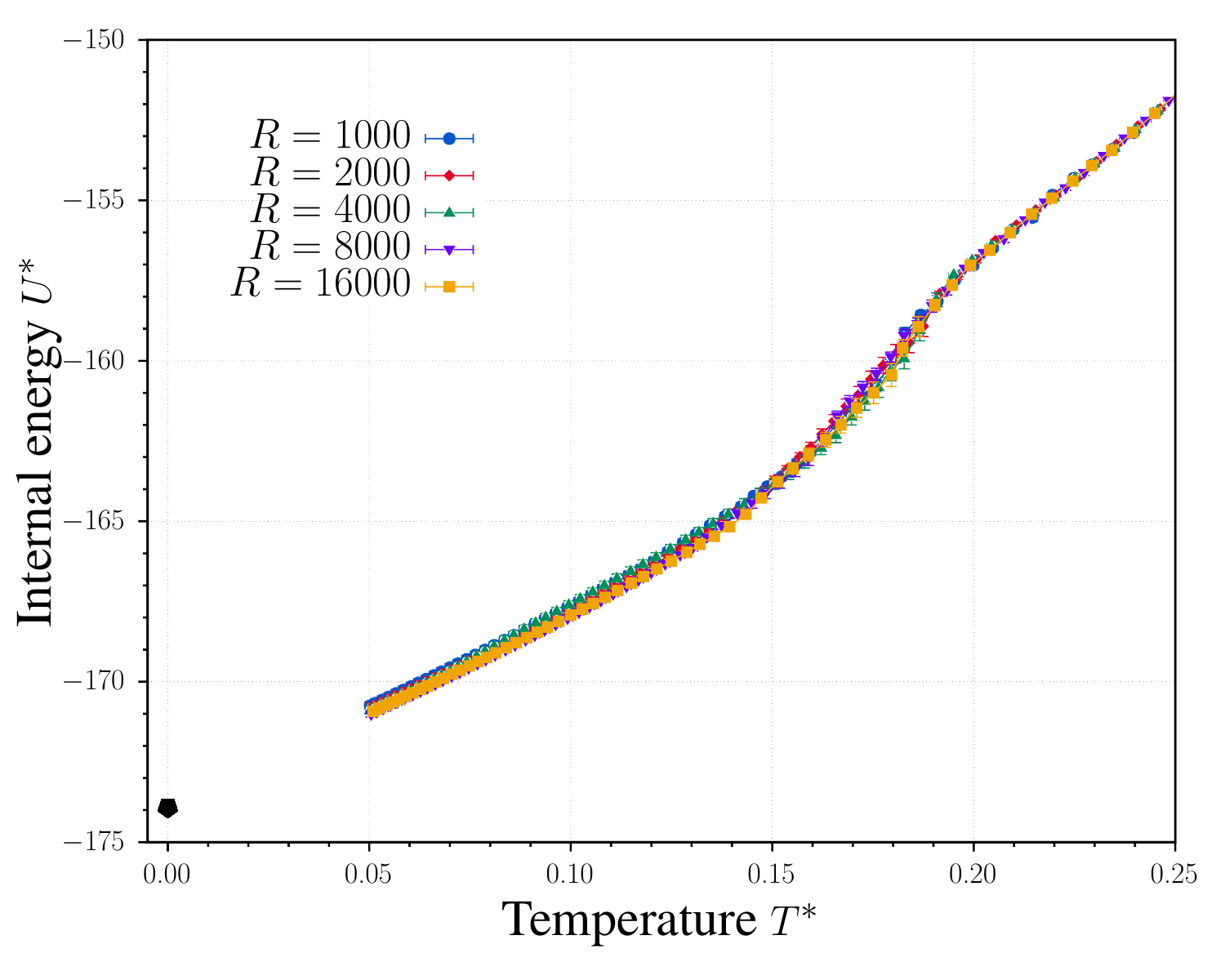}
\caption{Reduced internal energy $U^*(T^*)$ as a function of temperature $T^*$ for various population sizes $R$. Data points represent averages over $10$ independent PA runs, with error bars indicating the standard error. The consistent overlap of the curves, despite the adaptive temperature schedule being determined independently for each $R$, demonstrates the robustness of the sampling procedure. The symbol at $T^*=0$ denotes the global minimum energy reported in literature $(\simeq -173.92)$~\cite{wales1997global}. Our finite-temperature estimates extrapolate linearly toward this ground-state value as $T^*\to 0$. The pronounced change in slope at $T^*\simeq 0.17$ reflects the structural transition of the LJ$_{38}$ cluster. }
\label{energy}
\end{figure}

Fig.~\ref{energy} shows the reduced internal energy $U^*$ as a function of the reduced temperature $T^*$ for various population sizes.
For all $R$, the energy curves collapse within statistical error. Since the adaptive temperature schedule was independently determined for each population size to satisfy the ESS criterion, this consistency strongly suggests that the PA simulation reliably maintains the equilibrium ensemble throughout the cooling process. 
A pronounced change in the slope of $U^*(T^*)$ is observed around $T^*\simeq 0.17$. In the context of finite clusters like LJ$_{38}$, this feature corresponds to a melting-like structural transition. Unlike a sharp transition in the thermodynamic limit, this represents a thermal crossover where the system begins to fluctuate between low-energy solid-like structures and high-energy liquid-like states.

Importantly, the internal energy at low temperatures extrapolates linearly toward the known global minimum of $-173.92$~\cite{wales1997global}, as indicated by the mark at $T^*=0$. We further verified the sampling quality by quenching configurations sampled below the crossover temperature using the FIRE algorithm; these configurations consistently relaxed to the global minimum. This result confirms that our PA implementation effectively traverses the competitive energy landscape of LJ$_{38}$ and correctly populates the truncated octahedral basin, which is believed to be the true ground-state motif but is notoriously challenging to sample due to its narrow funnel.  

\begin{figure}[t] 
\centering
\includegraphics[width=3in]{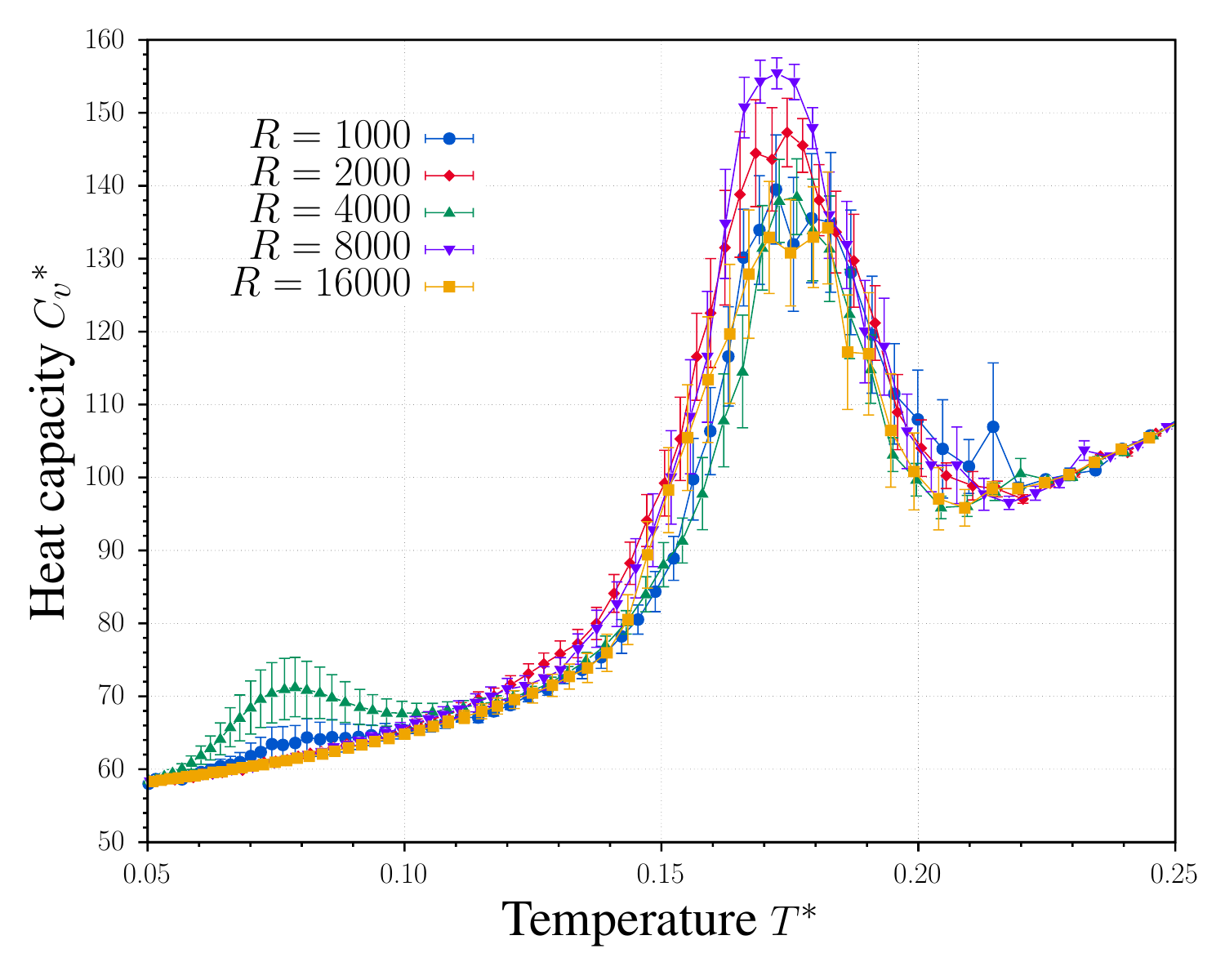}
\caption{Reduced heat capacity $C_v^*(T^*)(=C_v/k_{\mathrm B})$ estimated via PA for population sizes $R=1000$--$16000$. Error bars denote the standard error estimated from $10$ independent runs. The primary peak at $T^*\simeq 0.17$ signals the structural crossover of the LJ$_{38}$ cluster. The unstable feature observed at $T^*\simeq 0.08$ (most markedly for $R=4000$) is a numerical artifact arising from substantial inter-run fluctuations in the low-temperature regime. }
\label{heatcapacity}
\end{figure}

Fig.\ref{heatcapacity} presents temperature dependence of the reduced heat capacity $C_v^*(T^*)$, computed from the potential-energy fluctuations as $C_v^*\equiv C_v/k_{\mathrm B}=\beta^2(\langle E^2\rangle_{\beta}-\langle E\rangle_{\beta}^2)$. 
Consistent with the internal energy results, all population sizes $R$ exhibit a pronounced peak at $T^*\simeq 0.17$, characterizing the melting-like crossover between solid-like and liquid-like states. 
As $R$ increases, the peak becomes sharper, and the statistical uncertainty decreases, indicating improved statistical precision for larger populations.

A smaller secondary peak appears near $T^*\simeq 0.08$ for certain population sizes, specifically $R=4000$.  
This feature is not physically intrinsic and is interpreted as a numerical artifact resulting from incomplete equilibration. At these temperatures, the equilibrium state of the LJ$_{38}$ cluster is characterized by the coexistence of two competing structural funnels: the FCC-like and the icosahedral-like basins. Due to the high free energy barriers separating these motifs, independent runs exhibit large variations: some runs may become trapped in the icosahedral basin, while others successfully populate the FCC-like basin.
Since $C_v$ is proportional to the total energy variance of the sampled ensemble, a mixture of runs dominated by different basins artificially inflates the apparent fluctuations. This interpretation is supported by the large error bars in this temperature range, indicating that this feature is not robust across independent simulations. As $R$ increases, the probability that walkers discover the deep, narrow FCC-like basin increases. Once identified, the high-weight walkers are propagated through the population via resampling, yielding a more stable and representative sampling of the low-energy landscape. Consequently, these spurious peaks diminish as the population size becomes sufficient to represent the true equilibrium distribution. 

\subsection{Structural Analysis and Dimensionality Reduction}
\begin{figure}[t] 
\centering
\includegraphics[width=3.5in]{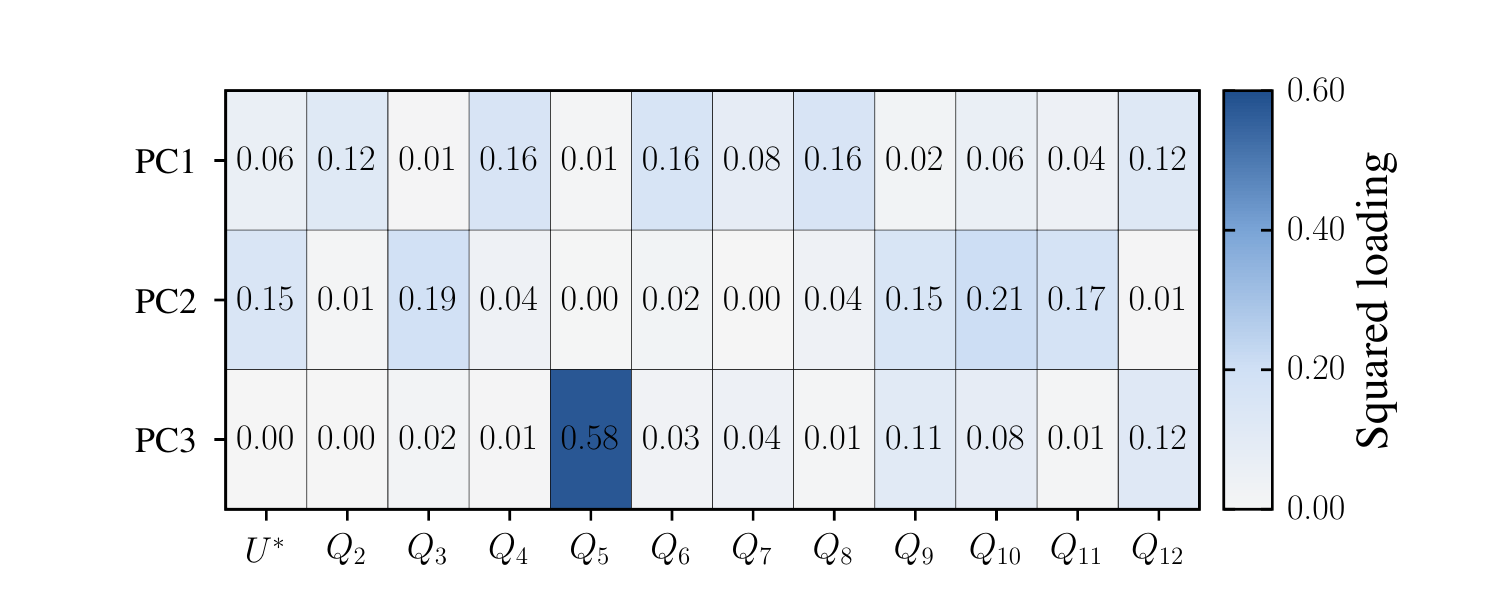}
\caption{
Heat map of squared loadings for the first three principal components (PC1--PC3). The PCA was performed on quenched configurations ($R=16000$), using a feature vector composed of the inherent-structure energy $U^*$ and the bond-orientational order parameters $Q_2$--$Q_{12}$ evaluated at the corresponding local minima. 
PC1, PC2, and PC3 explain 46.8\%, 14.7\%, and 10.9\% of the total variance, respectively.
}
\label{pca}
\end{figure}

As demonstrated above, the estimates of $U^*(T^*)$ and $C_v^*(T^*)$ become stable with respect to $R$ and the statistical uncertainties are reduced for large population sizes. 
On this basis, we fix the population size to $R=16000$ for the remainder of this study. To obtain structure-resolved thermodynamics, we analyze the quenched configurations collected at every temperature point throughout the entire cooling schedule from $10$ independent PA runs. This ensures that our structural classification captures the full transition from high-temperature liquid-like states to low-temperature crystalline-like motifs within a unified descriptor space. 

Using the $R=16000$ dataset, we examine which structural descriptors contribute most significantly to distinguishing the structural motifs.
We apply PCA to the standardized feature vectors, where each vector represents a single quenched configuration defined by its inherent-structure energy $U^*$ and the bond-orientational order parameters $\{Q_2,\dots, Q_{12}\}$ evaluated at the associated local minimum. Fig.\ref{pca} shows a heat map of the squared loadings for the first three principal components (PC1--PC3), where darker colors denote larger contributions to the variance captured by each component.
The explained-variance ratios of PC1, PC2, and PC3 are 46.8\%, 14.7\%, and 10.9\%, respectively, and thus these three components together account for approximately 72\% of the total variance.

The loading analysis in Fig.\ref{pca} reveals distinct roles for the various descriptors. 
PC1 is dominated by the even-$l$ bond-orientational order parameters $Q_4$, $Q_6$, and $Q_8$, indicating that the leading variance is primarily associated with the degree of bond-orientational order.
Notably, the contribution of $Q_8$ to PC1 is comparable to those of the conventional $Q_4$ and $Q_6$, suggesting that higher-order bond-orientational information provides complementary insight into the dominant structural variability of LJ$_{38}$.

\begin{figure}
    \centering
    \includegraphics[width=0.9\linewidth]{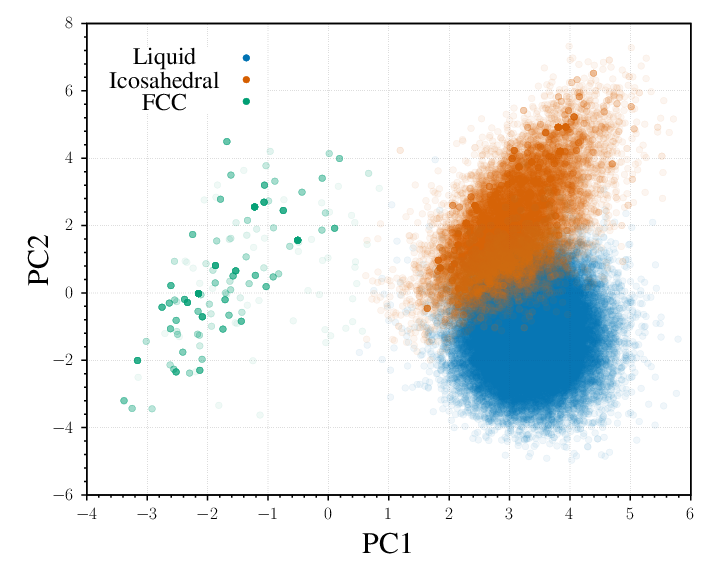}
    \caption{Projection plot of quenched configurations onto the first two principal components (PC1 and PC2). Each plot represents an inherent structure obtained via FIRE minimization, sampled from all temperature points throughout the PA simulations. Colors indicate the three structural clusters: liquid-like (blue), icosahedral-like (orange), and FCC-like (green), identified by $k$-means clustering in the PCA projection space. }
    \label{fig:PCA12}
\end{figure}
As shown in the PC1-PC2 projection in Fig.~\ref{fig:PCA12}, PC1 effectively separates the FCC-like basin from the non-FCC region containing both icosahedral and liquid-like configurations. PC2, which assigns high weights to descriptors with odd-$l$ values, such as $Q_3$ and $Q_{11}$, in addition to $U^*$ and $Q_{10}$, further distinguishes the icosahedral and liquid-like populations. The physical labels for these clusters were assigned a posteriori based on the characteristic inherent-structure energy and bond-orientational order observed within each group. This identification is consistent with the well-established energy landscape of LJ$_{38}$, which is dominated by the competition between the FCC truncated octahedral global minimum and the icosahedral funnel. Specifically, the cluster exhibiting the lowest $U^*$ and the highest crystallinity $(Q_4, Q_6, Q_8)$ is identified as the FCC-like basin, while the one with intermediate energy and order corresponds to the icosahedral family. The remaining cluster, characterized by high potential energy and negligible local order, represents the liquid-like state. 
This visualization confirms that the inherent structures of LJ$_{38}$ occupy well-defined regions in the reduced projection space, which enables a stable assignment of structural families.

\begin{figure}[t] 
\centering
\includegraphics[width=\linewidth]{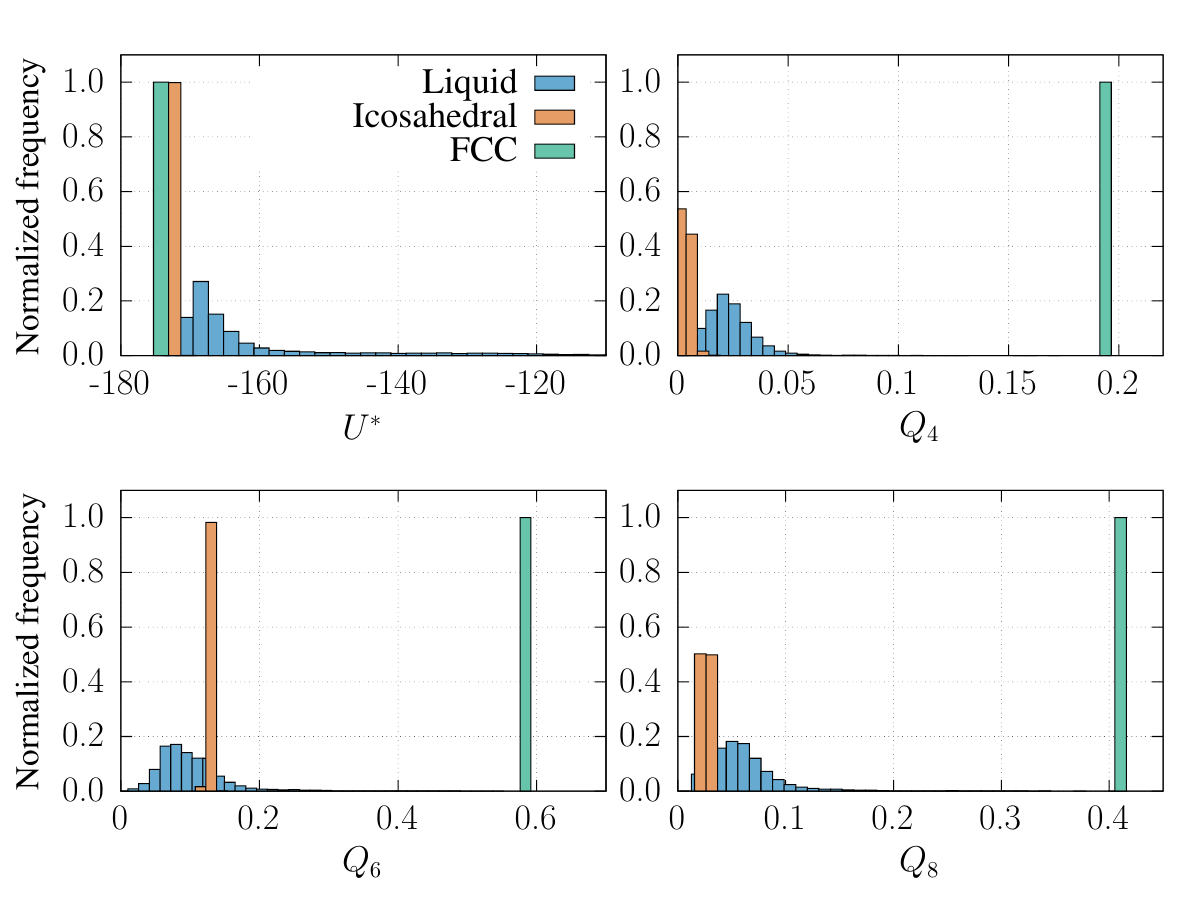}
\caption{
Normalized histograms of quenched configurations in terms of $U^\ast$ and $Q_4$, $Q_6$, and $Q_8$. The distributions are shown separately for the three structural clusters (liquid-like, icosahedral, FCC-like) identified via $k$-means clustering from the $R=16000$ PA runs.}
\label{histogram}
\end{figure}

Fig.~\ref{histogram} displays the normalized histograms of $U^*$, $Q_4$, $Q_6$, and $Q_8$ for the three structural clusters obtained by $k$-means clustering.
As noted above, these clusters are labeled liquid-like, icosahedral, and FCC-like based on their characteristic distributions in this descriptor space. The histograms show that the three families are well-separated, validating the robustness of the classification. 

For the inherent-structure energy $U^*$, the liquid-like cluster spans a broad range at higher values, reflecting an ensemble of diverse high-energy local minima. In contrast, the icosahedral and FCC-like clusters exhibit narrow, well-defined peaks at lower energies. Consistent with its identity as the global minimum, the FCC-like peak is located at a slightly lower energy than the icosahedral peak. The bond-orientational order parameters further clarify the structural distinctions. The FCC-like cluster exhibits high-intensity, narrow distributions at relatively large values of $Q_4, Q_6$,  and $Q_8$, reflecting the high degree of symmetry inherent in the truncated octahedral motif. The icosahedral cluster shows intermediate values that are clearly distinct from the FCC-like distributions. Conversely, the liquid-like cluster is characterized by broad distributions at low values, corresponding to weak local order of disordered configurations. The results demonstrate that
$k$-means clustering protocol successfully extracts three physically meaningful structural families from the quenched configurations. This provides a reliable labeling for the following free energy analysis.

\subsection{Free Energies of Structural Motif}
\begin{figure}[t] 
\centering
\includegraphics[width=3in]{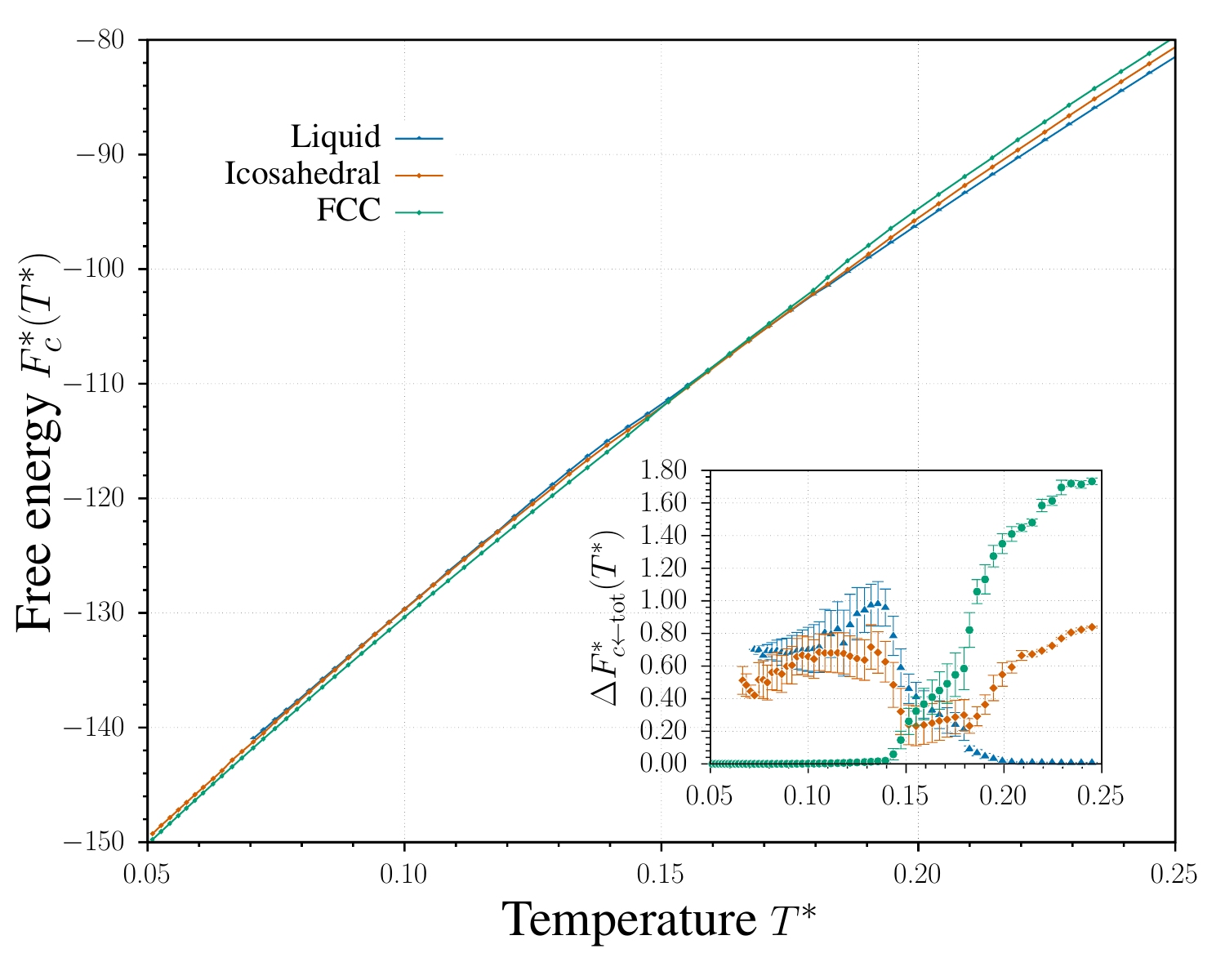}
\caption{
Main: Structure-resolved free energies $F_c^*(T^*)$ for the liquid-like, icosahedral, and FCC-like structures as a function of temperature. All values are referenced to the total ensemble free energy at the high-temperature, $F_{\rm tot}^*(T^*=1.0)=0$, where $F_\text{tot}^*(T^*)$ is computed from PA reweighting and normalization factors using successive partition-function ratios. 
Inset: free energy differences relative to the total ensemble, $\Delta F^*_{c\leftarrow \text{tot}}(T^*) = F_c^*(T^*)-F_{\rm tot}^*(T^*)=-T^*\ln(R_{c}/R)$, which quantify the statistical weight of each structural family $c$ within the population.
Results are obtained from 10 independent PA runs with R=16000, with points showing the run-averaged mean and error bars indicating the standard error across runs.
}
\label{deltaf}
\end{figure}

The thermodynamic stability of the identified structural motifs is quantified through their structure-resolved free energies. As shown in Fig.~\ref{deltaf}, these values are reconstructed by combining the temperature-dependent change in the total ensemble free energy with the population statistics of each basin (structural family) $c$. First, the total free energy change $F_\text{tot}^*(T^*)$ along the annealing process is computed from the PA reweighting and normalization factors, using the liquid-dominated state at $T^*=1.0$ as the reference. Subsequently, this ensemble free energy is partitioned into structure-resolved component $F_{c}^*(T^*)$ based on the population fractions of the classified walkers, $R_{c}/R$, for each family $c$. As described in Sec.~\ref{subsec:free_energy}, the use of population ratio is justified because the resampling procedure resets the walker weights to unity. The structure-resolved free energy is thus defined as $F_c^*(T^*) = F_\text{tot}^*(T^*) - T^*\ln(R_c/R)$. The inset of Fig.~\ref{deltaf} displays the term $-T^*\ln(R_c/R)$, representing the free energy contribution of each structural family relative to the total ensemble.

Because these structure-resolved estimates rely on the statistical distribution of walkers across the basins, the accuracy of the resulting free energy profiles depends strongly on the population size. The use of a large population ($R=16000$) is therefore essential to achieve sufficient resolution of population ratios, particularly in temperature regions where certain basins have low statistical weight.  At low $T^*$, minority basins may be represented by only a few walkers or even a single walker in the population, causing the corresponding low-temperature tail of $\Delta F_c^*(T^*)$ to be dominated by finite-population effects.

The resulting free energy profiles reveal two distinct crossovers determining the structural stability of the clusters. The first crossover, between the FCC-like and the icosahedral basins, occurs near $T^*\simeq 0.15$. Although the FCC-like truncated octahedron is the global energy minimum, the icosahedral basin becomes thermodynamically favored over the FCC-like basin upon heating. This crossover is driven by the high entropy of the icosahedral funnel, which contains numerous nearly degenerate local minima compared to the narrow FCC-like funnel. 

At higher temperatures, the liquid-like basin becomes increasingly competitive and eventually favored relative to the icosahedral basin.
The second crossover, between the icosahedral and liquid-like states, occurs around $T^*\simeq 0.17$. This temperature is in good agreement with the sharp peak in the heat capacity $C_v^*(T^*)$, as shown in Fig.~\ref{heatcapacity}, confirming that the primary thermodynamic feature of LJ$_{38}$ is indeed the transition into a disordered liquid-like phase. This ability to extract these detailed structure-resolved free energies from the population statistics of a single PA run demonstrates the significant practical advantage of this framework for exploring systems with multiple competing basins.

\section{Conclusion \& Future Work}
\label{sec:conclusion}
In this work, we studied the thermodynamic properties and structural transitions of the LJ$_{38}$ cluster. This system is widely recognized as a challenging benchmark due to its double-funnel energy landscape.
Using Population Annealing (PA) with an adaptive temperature schedule targeting $\mathrm{ESS}_{i+1}/R=0.95$, we demonstrated that thermodynamic observables, such as the reduced potential energy $U^*(T^*)$, and the reduced heat capacity $C_v^*(T^*)$, stably converge when the population size $R$ is sufficiently large. Specifically, our results for $R=16000$ provided well-converged estimates, whereas smaller populations exhibited incomplete equilibration between the competing FCC-like and icosahedral funnels, leading to numerical artifacts such as an apparent secondary feature in $C_v^*(T^*)$. These results emphasize the necessity of large-scale ensemble sampling to ensure the true equilibrium state in such frustrated systems.

The key contribution of this work is the integration of PA-based free energy estimation with a systematic structure-resolved analysis framework. By quenching sampled configurations to their nearest local minima, we effectively removed thermal noise, enabling a clear classification of inherent structures based on their fingerprints in the descriptor space spanned by $U^*$ and $Q_2$--$Q_{12}$. The application of PCA and $k$-means clustering to these quenched states provided an objective way to partition the complex configuration space into three physically interpretable structural families: FCC-like, icosahedral, and liquid-like.  

A significant advantage of this combined approach is its ability to correctly account for finite-temperature effects by directly sampling the equilibrium distribution. In contrast to conventional approaches that estimate free energies by adding harmonic or anharmonic contributions to zero-temperature potential energy of each structural basin, our method properly evaluates the statistical weight of paths from high-temperature configurations in the configuration space. Consequently, the free energy contribution of each structural basin is naturally partitioned through the population statistics. This allows for a robust inclusion of the configurational and vibrational entropy for each funnel without relying on local approximations or a priori assumptions about the basin's shape. Using this approach, we successfully quantified the thermodynamic competition between basins through the reduced free energy differences $\Delta F^*(T^*)$, locating the FCC--icosahedral crossover near $T^*\simeq 0.15$ and the icosahedral--liquid-like crossover around $T^*\simeq 0.17$. The latter is in excellent agreement with the sharp heat-capacity peak, confirming the transition into a disordered liquid.

While the present study set the number of clusters to $k=3$ based on the known structural motifs of LJ$_{38}$, a natural extension of this work is to incorporate automated clustering procedures to identify relevant structural families without prior assumptions. Such an extension would make the workflow applicable to more complex systems, including larger clusters or molecular models with richer competing motifs. Since PA is well-suited to massive parallelism, it offers a promising route toward structure-resolved thermodynamics on modern supercomputing architectures, providing an alternative sampling approach in molecular simulation.


\section*{Acknowledgment}
This work was supported by JSPS KAKENHI Grant Number JP25KJ0082, 23H01095, and JST Grant Number JPMJPF2221. The computations in this work were carried out using the facilities of the Supercomputer Center at the Institute for Solid State Physics, the University of Tokyo (ISSPkyodo-
SC-2024-Cb-0041, 2025-Ca-0081).


\bibliographystyle{IEEEtran} 
\bibliography{IEEEabrv}

\begin{thebibliography}{10}
\providecommand{\url}[1]{#1}
\csname url@samestyle\endcsname
\providecommand{\newblock}{\relax}
\providecommand{\bibinfo}[2]{#2}
\providecommand{\BIBentrySTDinterwordspacing}{\spaceskip=0pt\relax}
\providecommand{\BIBentryALTinterwordstretchfactor}{4}
\providecommand{\BIBentryALTinterwordspacing}{\spaceskip=\fontdimen2\font plus
\BIBentryALTinterwordstretchfactor\fontdimen3\font minus
  \fontdimen4\font\relax}
\providecommand{\BIBforeignlanguage}[2]{{%
\expandafter\ifx\csname l@#1\endcsname\relax
\typeout{** WARNING: IEEEtran.bst: No hyphenation pattern has been}%
\typeout{** loaded for the language `#1'. Using the pattern for}%
\typeout{** the default language instead.}%
\else
\language=\csname l@#1\endcsname
\fi
#2}}
\providecommand{\BIBdecl}{\relax}
\BIBdecl

\bibitem{hukushima2003population}
\BIBentryALTinterwordspacing
Hukushima,~K. and Iba,~Y., ``Population annealing and its application to a spin
  glass,'' \emph{AIP Conference Proceedings}, vol. 690, no.~1, pp. 200--206, 11
  2003. [Online]. Available: \url{https://doi.org/10.1063/1.1632130}
\BIBentrySTDinterwordspacing

\bibitem{Wang_2015}
\BIBentryALTinterwordspacing
Wang,~W., Machta,~J., and Katzgraber,~H.~G., ``Population annealing: Theory and
  application in spin glasses,'' \emph{Phys. Rev. E}, vol.~92, p. 063307, Dec
  2015. [Online]. Available:
  \url{https://link.aps.org/doi/10.1103/PhysRevE.92.063307}
\BIBentrySTDinterwordspacing

\bibitem{Callaham_2017}
\BIBentryALTinterwordspacing
Callaham,~J. and Machta,~J., ``Population annealing simulations of a binary
  hard-sphere mixture,'' \emph{Phys. Rev. E}, vol.~95, p. 063315, Jun 2017.
  [Online]. Available:
  \url{https://link.aps.org/doi/10.1103/PhysRevE.95.063315}
\BIBentrySTDinterwordspacing

\bibitem{Christiansen_2019PRL}
\BIBentryALTinterwordspacing
Christiansen,~H., Weigel,~M., and Janke,~W., ``Accelerating molecular dynamics
  simulations with population annealing,'' \emph{Phys. Rev. Lett.}, vol. 122,
  p. 060602, Feb 2019. [Online]. Available:
  \url{https://link.aps.org/doi/10.1103/PhysRevLett.122.060602}
\BIBentrySTDinterwordspacing

\bibitem{Machta_2010}
\BIBentryALTinterwordspacing
Machta,~J., ``Population annealing with weighted averages: A monte carlo method
  for rough free-energy landscapes,'' \emph{Phys. Rev. E}, vol.~82, p. 026704,
  Aug 2010. [Online]. Available:
  \url{https://link.aps.org/doi/10.1103/PhysRevE.82.026704}
\BIBentrySTDinterwordspacing

\bibitem{Wang_2015_Comp}
\BIBentryALTinterwordspacing
Wang,~W., Machta,~J., and Katzgraber,~H.~G., ``Comparing monte carlo methods
  for finding ground states of ising spin glasses: Population annealing,
  simulated annealing, and parallel tempering,'' \emph{Phys. Rev. E}, vol.~92,
  p. 013303, Jul 2015. [Online]. Available:
  \url{https://link.aps.org/doi/10.1103/PhysRevE.92.013303}
\BIBentrySTDinterwordspacing

\bibitem{hukushima1996exchange}
\BIBentryALTinterwordspacing
Hukushima,~K. and Nemoto,~K., ``Exchange monte carlo method and application to
  spin glass simulations,'' \emph{Journal of the Physical Society of Japan},
  vol.~65, no.~6, pp. 1604--1608, 1996. [Online]. Available:
  \url{https://doi.org/10.1143/JPSJ.65.1604}
\BIBentrySTDinterwordspacing

\bibitem{Weigel_2017}
\BIBentryALTinterwordspacing
Weigel,~M., Barash,~L.~Y., Borovský,~M., Janke,~W., and Shchur,~L.~N.,
  ``Population annealing: Massively parallel simulations in statistical
  physics,'' \emph{Journal of Physics: Conference Series}, vol. 921, no.~1, p.
  012017, nov 2017. [Online]. Available:
  \url{https://doi.org/10.1088/1742-6596/921/1/012017}
\BIBentrySTDinterwordspacing

\bibitem{BARASH2017341}
\BIBentryALTinterwordspacing
Barash,~L.~Y., Weigel,~M., Borovský,~M., Janke,~W., and Shchur,~L.~N., ``Gpu
  accelerated population annealing algorithm,'' \emph{Computer Physics
  Communications}, vol. 220, pp. 341--350, 2017. [Online]. Available:
  \url{https://www.sciencedirect.com/science/article/pii/S0010465517302023}
\BIBentrySTDinterwordspacing

\bibitem{shechtman1984metallic}
\BIBentryALTinterwordspacing
Shechtman,~D., Blech,~I., Gratias,~D., and Cahn,~J.~W., ``Metallic phase with
  long-range orientational order and no translational symmetry,'' \emph{Phys.
  Rev. Lett.}, vol.~53, pp. 1951--1953, Nov 1984. [Online]. Available:
  \url{https://link.aps.org/doi/10.1103/PhysRevLett.53.1951}
\BIBentrySTDinterwordspacing

\bibitem{tsai1987stable}
\BIBentryALTinterwordspacing
Tsai,~A.-P., Inoue,~A., and Masumoto,~T., ``A stable quasicrystal in al-cu-fe
  system,'' \emph{Japanese Journal of Applied Physics}, vol.~26, no.~9A, p.
  L1505, sep 1987. [Online]. Available:
  \url{https://doi.org/10.1143/JJAP.26.L1505}
\BIBentrySTDinterwordspacing

\bibitem{ebert}
\BIBentryALTinterwordspacing
Ebert,~P., Feuerbacher,~M., Tamura,~N., Wollgarten,~M., and Urban,~K.,
  ``Evidence for a cluster-based structure of alpdmn single quasicrystals,''
  \emph{Phys. Rev. Lett.}, vol.~77, pp. 3827--3830, Oct 1996. [Online].
  Available: \url{https://link.aps.org/doi/10.1103/PhysRevLett.77.3827}
\BIBentrySTDinterwordspacing

\bibitem{marks1981425}
\BIBentryALTinterwordspacing
Marks,~L. and Smith,~D.~J., ``High resolution studies of small particles of
  gold and silver: I. multiply-twinned particles,'' \emph{Journal of Crystal
  Growth}, vol.~54, no.~3, pp. 425--432, 1981. [Online]. Available:
  \url{https://www.sciencedirect.com/science/article/pii/0022024881904942}
\BIBentrySTDinterwordspacing

\bibitem{vanderVelden1981}
\BIBentryALTinterwordspacing
van~der Velden,~J. W.~A., Bour,~J.~J., Bosman,~W.~P., and Noordik,~J.~H.,
  ``Synthesis and x-ray crystal structure determination of the cationic gold
  cluster compound [au8(pph3)7](no3)2,'' \emph{J. Chem. Soc.{,} Chem. Commun.},
  pp. 1218--1219, 1981. [Online]. Available:
  \url{http://dx.doi.org/10.1039/C39810001218}
\BIBentrySTDinterwordspacing

\bibitem{fan2010self}
\BIBentryALTinterwordspacing
Fan,~J.~A., Wu,~C., Bao,~K., Bao,~J., Bardhan,~R., Halas,~N.~J.,
  Manoharan,~V.~N., Nordlander,~P., Shvets,~G., and Capasso,~F.,
  ``Self-assembled plasmonic nanoparticle clusters,'' \emph{Science}, vol. 328,
  no. 5982, pp. 1135--1138, 2010. [Online]. Available:
  \url{https://www.science.org/doi/abs/10.1126/science.1187949}
\BIBentrySTDinterwordspacing

\bibitem{de2015entropy}
\BIBentryALTinterwordspacing
De~Nijs,~B., Dussi,~S., Smallenburg,~F., Meeldijk,~J.~D., Groenendijk,~D.~J.,
  Filion,~L., Imhof,~A., Van~Blaaderen,~A., and Dijkstra,~M., ``Entropy-driven
  formation of large icosahedral colloidal clusters by spherical confinement,''
  \emph{Nature materials}, vol.~14, no.~1, pp. 56--60, 2015. [Online].
  Available: \url{https://doi.org/10.1038/nmat4072}
\BIBentrySTDinterwordspacing

\bibitem{noya2006structural}
\BIBentryALTinterwordspacing
Noya,~E.~G. and Doye,~J. P.~K., ``Structural transitions in the 309-atom magic
  number lennard-jones cluster,'' \emph{The Journal of Chemical Physics}, vol.
  124, no.~10, p. 104503, 03 2006. [Online]. Available:
  \url{https://doi.org/10.1063/1.2173260}
\BIBentrySTDinterwordspacing

\bibitem{doye1999evolution}
\BIBentryALTinterwordspacing
Doye,~J. P.~K., Miller,~M.~A., and Wales,~D.~J., ``Evolution of the potential
  energy surface with size for lennard-jones clusters,'' \emph{The Journal of
  Chemical Physics}, vol. 111, no.~18, pp. 8417--8428, 11 1999. [Online].
  Available: \url{https://doi.org/10.1063/1.480217}
\BIBentrySTDinterwordspacing

\bibitem{doye2002entropic}
\BIBentryALTinterwordspacing
Doye,~J. P.~K. and Calvo,~F., ``Entropic effects on the structure of
  lennard-jones clusters,'' \emph{The Journal of Chemical Physics}, vol. 116,
  no.~19, pp. 8307--8317, 05 2002. [Online]. Available:
  \url{https://doi.org/10.1063/1.1469616}
\BIBentrySTDinterwordspacing

\bibitem{northby1987structure}
\BIBentryALTinterwordspacing
Northby,~J.~A., ``Structure and binding of lennard-jones clusters: $13\geq
  n\geq 147$,'' \emph{The Journal of Chemical Physics}, vol.~87, no.~10, pp.
  6166--6177, 11 1987. [Online]. Available:
  \url{https://doi.org/10.1063/1.453492}
\BIBentrySTDinterwordspacing

\bibitem{sehgal2014phase}
\BIBentryALTinterwordspacing
Sehgal,~R.~M., Maroudas,~D., and Ford,~D.~M., ``Phase behavior of the 38-atom
  lennard-jones cluster,'' \emph{The Journal of Chemical Physics}, vol. 140,
  no.~10, p. 104312, 03 2014. [Online]. Available:
  \url{https://doi.org/10.1063/1.4866810}
\BIBentrySTDinterwordspacing

\bibitem{doye1995effect}
\BIBentryALTinterwordspacing
Doye,~J. P.~K., Wales,~D.~J., and Berry,~R.~S., ``The effect of the range of
  the potential on the structures of clusters,'' \emph{The Journal of Chemical
  Physics}, vol. 103, no.~10, pp. 4234--4249, 09 1995. [Online]. Available:
  \url{https://doi.org/10.1063/1.470729}
\BIBentrySTDinterwordspacing

\bibitem{doye1997structural}
\BIBentryALTinterwordspacing
P.~K.~Doye,~J. and J.~Wales,~D., ``Structural consequences of the range of the
  interatomic potential a menagerie of clusters,'' \emph{J. Chem. Soc.{,}
  Faraday Trans.}, vol.~93, pp. 4233--4243, 1997. [Online]. Available:
  \url{http://dx.doi.org/10.1039/A706221D}
\BIBentrySTDinterwordspacing

\bibitem{doye1999double}
\BIBentryALTinterwordspacing
Doye,~J. P.~K., Miller,~M.~A., and Wales,~D.~J., ``The double-funnel energy
  landscape of the 38-atom lennard-jones cluster,'' \emph{The Journal of
  Chemical Physics}, vol. 110, no.~14, pp. 6896--6906, 04 1999. [Online].
  Available: \url{https://doi.org/10.1063/1.478595}
\BIBentrySTDinterwordspacing

\bibitem{partay2010efficient}
\BIBentryALTinterwordspacing
Pártay,~L.~B., Bartók,~A.~P., and Csányi,~G., ``Efficient sampling of atomic
  configurational spaces,'' \emph{The Journal of Physical Chemistry B}, vol.
  114, no.~32, pp. 10\,502--10\,512, 2010, pMID: 20701382. [Online]. Available:
  \url{https://doi.org/10.1021/jp1012973}
\BIBentrySTDinterwordspacing

\bibitem{poulain2006performances}
\BIBentryALTinterwordspacing
Poulain,~P., Calvo,~F., Antoine,~R., Broyer,~M., and Dugourd,~P.,
  ``Performances of wang-landau algorithms for continuous systems,''
  \emph{Phys. Rev. E}, vol.~73, p. 056704, May 2006. [Online]. Available:
  \url{https://link.aps.org/doi/10.1103/PhysRevE.73.056704}
\BIBentrySTDinterwordspacing

\bibitem{bogdan2006equilibrium}
Bogdan,~T.~V., Wales,~D.~J., and Calvo,~F., ``Equilibrium thermodynamics from
  basin-sampling,'' \emph{The Journal of Chemical Physics}, vol. 124, no.~4, p.
  044102, 01 2006.

\bibitem{lv2012particle}
Lv,~J., Wang,~Y., Zhu,~L., and Ma,~Y., ``Particle-swarm structure prediction on
  clusters,'' \emph{The Journal of Chemical Physics}, vol. 137, no.~8, p.
  084104, 08 2012.

\bibitem{goedecker2004minima}
\BIBentryALTinterwordspacing
Goedecker,~S., ``Minima hopping: An efficient search method for the global
  minimum of the potential energy surface of complex molecular systems,''
  \emph{The Journal of Chemical Physics}, vol. 120, no.~21, pp. 9911--9917, 06
  2004. [Online]. Available: \url{https://doi.org/10.1063/1.1724816}
\BIBentrySTDinterwordspacing

\bibitem{wales2006potential}
\BIBentryALTinterwordspacing
Wales,~D.~J. and Bogdan,~T.~V., ``Potential energy and free energy
  landscapes,'' \emph{The Journal of Physical Chemistry B}, vol. 110, no.~42,
  pp. 20\,765--20\,776, 2006, pMID: 17048885. [Online]. Available:
  \url{https://doi.org/10.1021/jp0680544}
\BIBentrySTDinterwordspacing

\bibitem{neirotti2000phase}
\BIBentryALTinterwordspacing
Neirotti,~J.~P., Calvo,~F., Freeman,~D.~L., and Doll,~J.~D., ``Phase changes in
  38-atom lennard-jones clusters. i. a parallel tempering study in the
  canonical ensemble,'' \emph{The Journal of Chemical Physics}, vol. 112,
  no.~23, pp. 10\,340--10\,349, 06 2000. [Online]. Available:
  \url{https://doi.org/10.1063/1.481671}
\BIBentrySTDinterwordspacing

\bibitem{avendano2016firefly}
\BIBentryALTinterwordspacing
Avenda{\~n}o-Franco,~G. and Romero,~A.~H., ``Firefly algorithm for structural
  search,'' \emph{Journal of Chemical Theory and Computation}, vol.~12, no.~7,
  pp. 3416--3428, 2016, pMID: 27232694. [Online]. Available:
  \url{https://doi.org/10.1021/acs.jctc.5b01157}
\BIBentrySTDinterwordspacing

\bibitem{predescu2005thermodynamics}
\BIBentryALTinterwordspacing
Predescu,~C., Frantsuzov,~P.~A., and Mandelshtam,~V.~A., ``Thermodynamics and
  equilibrium structure of ne38 cluster: Quantum mechanics versus classical,''
  \emph{The Journal of Chemical Physics}, vol. 122, no.~15, p. 154305, 04 2005.
  [Online]. Available: \url{https://doi.org/10.1063/1.1860331}
\BIBentrySTDinterwordspacing

\bibitem{schaefer2014minima}
\BIBentryALTinterwordspacing
Schaefer,~B., Mohr,~S., Amsler,~M., and Goedecker,~S., ``Minima hopping guided
  path search: An efficient method for finding complex chemical reaction
  pathways,'' \emph{The Journal of Chemical Physics}, vol. 140, no.~21, p.
  214102, 06 2014. [Online]. Available: \url{https://doi.org/10.1063/1.4878944}
\BIBentrySTDinterwordspacing

\bibitem{kanayama2023structure}
\BIBentryALTinterwordspacing
Kanayama,~K., Seko,~A., and Toyoura,~K., ``Structure search method for atomic
  clusters based on the dividing rectangles algorithm,'' \emph{Phys. Rev. E},
  vol. 108, p. 035303, Sep 2023. [Online]. Available:
  \url{https://link.aps.org/doi/10.1103/PhysRevE.108.035303}
\BIBentrySTDinterwordspacing

\bibitem{sharapov2007solid}
\BIBentryALTinterwordspacing
Sharapov,~V.~A. and Mandelshtam,~V.~A., ``Solid-solid structural
  transformations in lennard-jones clusters: Accurate simulations versus the
  harmonic superposition approximation,'' \emph{The Journal of Physical
  Chemistry A}, vol. 111, no.~41, pp. 10\,284--10\,291, 2007, pMID: 17685597.
  [Online]. Available: \url{https://doi.org/10.1021/jp072929c}
\BIBentrySTDinterwordspacing

\bibitem{liu2005convergence}
\BIBentryALTinterwordspacing
Liu,~H. and Jordan,~K.~D., ``On the convergence of parallel tempering monte
  carlo simulations of lj38,'' \emph{The Journal of Physical Chemistry A}, vol.
  109, no.~23, pp. 5203--5207, 2005, pMID: 16833877. [Online]. Available:
  \url{https://doi.org/10.1021/jp050367w}
\BIBentrySTDinterwordspacing

\bibitem{berg1991multicanonical}
\BIBentryALTinterwordspacing
Berg,~B.~A. and Neuhaus,~T., ``Multicanonical algorithms for first order phase
  transitions,'' \emph{Physics Letters B}, vol. 267, no.~2, pp. 249--253, 1991.
  [Online]. Available:
  \url{https://www.sciencedirect.com/science/article/pii/037026939191256U}
\BIBentrySTDinterwordspacing

\bibitem{berg1992multicanonical}
\BIBentryALTinterwordspacing
------, ``Multicanonical ensemble: A new approach to simulate first-order phase
  transitions,'' \emph{Phys. Rev. Lett.}, vol.~68, pp. 9--12, Jan 1992.
  [Online]. Available: \url{https://link.aps.org/doi/10.1103/PhysRevLett.68.9}
\BIBentrySTDinterwordspacing

\bibitem{wang2001efficient}
\BIBentryALTinterwordspacing
Wang,~F. and Landau,~D.~P., ``Efficient, multiple-range random walk algorithm
  to calculate the density of states,'' \emph{Phys. Rev. Lett.}, vol.~86, pp.
  2050--2053, Mar 2001. [Online]. Available:
  \url{https://link.aps.org/doi/10.1103/PhysRevLett.86.2050}
\BIBentrySTDinterwordspacing

\bibitem{wales1997global}
\BIBentryALTinterwordspacing
Wales,~D.~J. and Doye,~J. P.~K., ``Global optimization by basin-hopping and the
  lowest energy structures of lennard-jones clusters containing up to 110
  atoms,'' \emph{The Journal of Physical Chemistry A}, vol. 101, no.~28, pp.
  5111--5116, 1997. [Online]. Available:
  \url{https://doi.org/10.1021/jp970984n}
\BIBentrySTDinterwordspacing

\bibitem{skilling2004nested}
\BIBentryALTinterwordspacing
Skilling,~J., ``{Nested sampling for general Bayesian computation},''
  \emph{Bayesian Analysis}, vol.~1, no.~4, pp. 833 -- 859, 2006. [Online].
  Available: \url{https://doi.org/10.1214/06-BA127}
\BIBentrySTDinterwordspacing

\bibitem{cezar2017parallel}
\BIBentryALTinterwordspacing
Cezar,~H.~M., Rondina,~G.~G., and Da~Silva,~J. L.~F., ``Parallel tempering
  monte carlo combined with clustering euclidean metric analysis to study the
  thermodynamic stability of lennard-jones nanoclusters,'' \emph{The Journal of
  Chemical Physics}, vol. 146, no.~6, p. 064114, 02 2017. [Online]. Available:
  \url{https://doi.org/10.1063/1.4975601}
\BIBentrySTDinterwordspacing

\bibitem{Doucet2001}
\BIBentryALTinterwordspacing
Doucet,~A., de~Freitas,~N., and Gordon,~N., \emph{An Introduction to Sequential
  Monte Carlo Methods}.\hskip 1em plus 0.5em minus 0.4em\relax New York, NY:
  Springer New York, 2001, pp. 3--14. [Online]. Available:
  \url{https://doi.org/10.1007/978-1-4757-3437-9_1}
\BIBentrySTDinterwordspacing

\bibitem{adaptive_resampling}
\BIBentryALTinterwordspacing
Moral,~P.~D., Doucet,~A., and Jasra,~A., ``On adaptive resampling strategies
  for sequential monte carlo methods,'' \emph{Bernoulli}, vol.~18, no.~1, pp.
  252--278, 2012. [Online]. Available:
  \url{http://www.jstor.org/stable/23238595}
\BIBentrySTDinterwordspacing

\bibitem{Tutorial_particle_filtering}
\BIBentryALTinterwordspacing
Doucet,~A. and Johansen,~A.~M., ``A tutorial on particle filtering and
  smoothing : fiteen years later,'' in \emph{The Oxford handbook of nonlinear
  filtering}, ser. Oxford handbooks in mathematics, Crisan,~D. and
  Rozovskii,~B., Eds.\hskip 1em plus 0.5em minus 0.4em\relax Oxford ; N.Y.:
  Oxford University Press, 2011, pp. 656--705. [Online]. Available:
  \url{http://webcat.warwick.ac.uk/record=b2490036~S1}
\BIBentrySTDinterwordspacing

\bibitem{gessert2023resampling}
\BIBentryALTinterwordspacing
Gessert,~D., Janke,~W., and Weigel,~M., ``Resampling schemes in population
  annealing: Numerical and theoretical results,'' \emph{Phys. Rev. E}, vol.
  108, p. 065309, Dec 2023. [Online]. Available:
  \url{https://link.aps.org/doi/10.1103/PhysRevE.108.065309}
\BIBentrySTDinterwordspacing

\bibitem{Christiansen_2019}
\BIBentryALTinterwordspacing
Christiansen,~H., Weigel,~M., and Janke,~W., ``Population annealing molecular
  dynamics with adaptive temperature steps,'' \emph{Journal of Physics:
  Conference Series}, vol. 1163, no.~1, p. 012074, feb 2019. [Online].
  Available: \url{https://doi.org/10.1088/1742-6596/1163/1/012074}
\BIBentrySTDinterwordspacing

\bibitem{SA1983}
\BIBentryALTinterwordspacing
Kirkpatrick,~S., Gelatt,~C.~D., and Vecchi,~M.~P., ``Optimization by simulated
  annealing,'' \emph{Science}, vol. 220, no. 4598, pp. 671--680, 1983.
  [Online]. Available:
  \url{https://www.science.org/doi/abs/10.1126/science.220.4598.671}
\BIBentrySTDinterwordspacing

\bibitem{Kirkpatrick1984}
\BIBentryALTinterwordspacing
Kirkpatrick,~S., ``Optimization by simulated annealing: Quantitative studies,''
  \emph{Journal of Statistical Physics}, vol.~34, no.~5, pp. 975--986, Mar
  1984. [Online]. Available: \url{https://doi.org/10.1007/BF01009452}
\BIBentrySTDinterwordspacing

\bibitem{Rossky_1978}
\BIBentryALTinterwordspacing
Rossky,~P.~J., Doll,~J.~D., and Friedman,~H.~L., ``Brownian dynamics as smart
  monte carlo simulation,'' \emph{The Journal of Chemical Physics}, vol.~69,
  no.~10, pp. 4628--4633, 11 1978. [Online]. Available:
  \url{https://doi.org/10.1063/1.436415}
\BIBentrySTDinterwordspacing

\bibitem{Roberts1996ExponentialCO}
\BIBentryALTinterwordspacing
Roberts,~G.~O. and Tweedie,~R.~L., ``Exponential convergence of langevin
  distributions and their discrete approximations,'' \emph{Bernoulli}, vol.~2,
  pp. 341--363, 1996. [Online]. Available:
  \url{https://api.semanticscholar.org/CorpusID:18787082}
\BIBentrySTDinterwordspacing

\bibitem{Roberts_2002}
\BIBentryALTinterwordspacing
Roberts,~G.~O. and Rosenthal,~J.~S., ``Optimal scaling of discrete
  approximations to langevin diffusions,'' \emph{Journal of the Royal
  Statistical Society Series B: Statistical Methodology}, vol.~60, no.~1, pp.
  255--268, 01 2002. [Online]. Available:
  \url{https://doi.org/10.1111/1467-9868.00123}
\BIBentrySTDinterwordspacing

\bibitem{steinhardt1983bond}
\BIBentryALTinterwordspacing
Steinhardt,~P.~J., Nelson,~D.~R., and Ronchetti,~M., ``Bond-orientational order
  in liquids and glasses,'' \emph{Phys. Rev. B}, vol.~28, pp. 784--805, Jul
  1983. [Online]. Available:
  \url{https://link.aps.org/doi/10.1103/PhysRevB.28.784}
\BIBentrySTDinterwordspacing

\bibitem{bitzek2006structural}
\BIBentryALTinterwordspacing
Bitzek,~E., Koskinen,~P., G\"ahler,~F., Moseler,~M., and Gumbsch,~P.,
  ``Structural relaxation made simple,'' \emph{Phys. Rev. Lett.}, vol.~97, p.
  170201, Oct 2006. [Online]. Available:
  \url{https://link.aps.org/doi/10.1103/PhysRevLett.97.170201}
\BIBentrySTDinterwordspacing

\bibitem{macqueen1967multivariate}
MacQueen,~J., ``Multivariate observations,'' in \emph{Proceedings ofthe 5th
  Berkeley symposium on mathematical statisticsand probability}, vol.~1.\hskip
  1em plus 0.5em minus 0.4em\relax University of California press Oakland, CA,
  USA, 1967, pp. 281--297.

\end{thebibliography}

\end{document}